\documentclass[aps]{revtex4}
\newcommand{\bb}{\begin{eqnarray}}
\newcommand{\ee}{\end{eqnarray}}
\begin{document}
\title{{Tunnelling from black holes and tunnelling into white holes}}
\author{Bhramar Chatterjee}\email{bhramar.chatterjee@saha.ac.in}
\author{A. Ghosh}\email{amit.ghosh@saha.ac.in}
\author{P. Mitra}\email{parthasarathi.mitra@saha.ac.in}
\affiliation{Saha Institute of Nuclear Physics\\
Block AF, Bidhannagar\\
Calcutta 700 064}
\begin{abstract}
Hawking radiation is nowadays being understood as tunnelling through black hole horizons.
Here, the extension of the Hamilton-Jacobi approach to tunnelling for
non-rotating and rotating black holes in different 
non-singular coordinate systems not only confirms this quantum 
emission from black holes but also reveals the new phenomenon 
of absorption into white holes by quantum mechanical tunnelling.
The r\^{o}le of a boundary condition of total absorption or
emission is also clarified.
\end{abstract}
\maketitle

\noindent {\bf 1.} A classical black hole has a horizon beyond which nothing
escapes to the outside world. But there is a relation connecting the area
of the horizon with the mass (and other parameters like the charge)
indicating a close similarity \cite{BCH} with
the thermodynamical laws, suggesting the introduction of an entropy
and a temperature \cite{Bek}. 
This analogy was interpreted to be of quantum origin and the scale fixed
after the (theoretical) discovery of radiation from black holes \cite{Hawk}. 
For the simplest Schwarzschild black hole, this radiation, which is thermal, has a
temperature $T_H={\hbar\over 4\pi r_h}={\hbar\over 8\pi M},$
where $r_h$ gives the location of the horizon in standard coordinates
and $M$ is the mass of the black hole.
This result was derived by considering quantum massless particles in a
Schwarzschild background geometry. 
A Hamilton - Jacobi equation based on a quantum mechanical tunnelling picture 
is more intuitive \cite{pad,zerbini,mann}, but special care is 
needed to reproduce the temperature \cite{mitra}, avoiding incorrect values.
Some other interesting tunnelling discussions may be found in \cite{others}.

A massless particle in a black hole background is described
by the Klein-Gordon equation
\bb
\hbar^2(-g)^{-1/2}\partial_\mu(g^{\mu\nu}(-g)^{1/2}\partial_\nu\phi)=0.
\ee
One expands
\bb
\phi=\exp(-{i\over\hbar}S+...)
\ee
to obtain to leading order in $\hbar$ the Hamilton-Jacobi equation
\bb
g^{\mu\nu}\partial_\mu S\partial_\nu S=0.
\ee
If the black hole horizon is a Killing horizon such that there is a Killing vector $\chi\equiv
{\partial\over\partial\tau}$ which is timelike in a vicinity of the horizon 
and becomes null on the horizon, a surface 
gravity $\kappa$ can be defined as follows:
\bb
\chi\cdot\nabla\chi^\mu=\kappa\chi^\mu\quad {\rm on~the~horizon}.
\ee
It is convenient to introduce a spatial coordinate $\lambda$ by
\bb
\chi^2=-2\kappa\lambda.
\ee
Then {\it near the horizon} the metric can be written in the form \cite{olafg}
\bb
ds^2=-2\kappa\lambda d\tau^2+(2\kappa\lambda)^{-1}d\lambda^2+{\rm other~terms}.
\ee
Separation of variables yields a solution independent of the
other (angular) spatial coordinates:
\bb
S=E\tau+C+S_0(\lambda),
\ee
where $E$ is to be interpreted as the energy and
$C$ is a constant. The equation for $S_0$ then reads
\bb
-{E^2\over 2\kappa\lambda} + (2\kappa\lambda)S_0'(\lambda)^2=0.
\ee
The formal solution of this equation is
\bb
S_0(\lambda)=\pm E\int^\lambda{d\lambda\over  2\kappa\lambda}.
\ee
The sign ambiguity comes from the square root and corresponds to the
fact that there can be incoming/outgoing solutions. There is, furthermore,
a singularity at the horizon $\lambda=0$, which has to be handled if one tries to find
a solution across it. 
One way to skirt this pole is to change $\lambda$ to  $\lambda-i\epsilon$
in the denominator. This yields
\bb
S_{out}&=&E\tau+C-{E\over 2\kappa}[i\pi+\int^\lambda d\lambda P(\frac{1}{\lambda})],\nonumber\\
S_{in}&=&E\tau+C+{E\over 2\kappa}[i\pi+\int^\lambda d\lambda P(\frac{1}{\lambda})],\label{ipi}
\ee
where $P()$ denotes the principal value. It was suggested in \cite{mitra} that 
the imaginary part of $C$
is to be determined so as to cancel the imaginary part
of $S_{in}$ and thus to ensure that the incoming probability
is unity {\it in the classical limit} -- when there is no reflection and everything is 
absorbed -- instead of zero or infinity. Thus,
\bb
C&=&-i\pi {E\over 2\kappa} + (Re~C),\nonumber\\
S_{out}&=&E\tau-{E\over 2\kappa}[2i\pi+\int^\lambda d\lambda P(\frac{1}{\lambda})]+ (Re~C),
\ee
implying a decay factor $\exp (-\pi E/\hbar\kappa)$ in the amplitude,
and a factor $\exp (-2\pi E/\hbar\kappa)$ in the probability,
in conformity with the standard value $\frac{\hbar\kappa}{2\pi}$ of the Hawking temperature.
Note that in the Schwarzschild case, $\lambda$ is $r-r_h$ and $\kappa=\frac{1}{2r_h}$.

However, the use of singular coordinates across horizons is not reliable,
and the procedure must be regarded at best as an ad hoc one.
Further, the introduction of an imaginary part in $C$
may be felt to be undesirable in a non-singular framework. 
In the following pages, we shall go through a
study of the non-rotating black hole as well as the Kerr black hole in several 
non-singular coordinate systems {\it without including an imaginary part in $C$}.
An analysis of the results in the concluding section
will show that not only can black holes radiate, even {\em
white holes are capable of the classically forbidden process of
absorption by quantum mechanical tunnelling}.
The need for an imaginary part of $C$ in the case of singular coordinate
systems will also become clear.
                                                                                                                       
\noindent{\bf 2.} Let us first consider the advanced Eddington-Finkelstein coordinates
to describe the Schwarzschild black hole. Here one employs the coordinate
$v$ instead of $t$, where
\bb
dv=dt+{dr\over 1-\frac{r_h}{r}}.
\ee
The metric reads
\bb
ds^2=-(1-\frac{r_h}{r})dv^2+ 2dvdr+r^2d\Omega^2.
\ee
The Hamilton-Jacobi equation for spherically symmetric $S$ is
\bb
2{\partial S\over \partial v}{\partial S\over \partial r}+
(1-\frac{r_h}{r})({\partial S\over \partial r})^2=0,
\ee
in which we can put ${\partial S\over \partial v}=E$ to obtain
two solutions
\bb
S_{out}=Ev-2E\int{dr\over 1-\frac{r_h}{r}},\quad
S_{in}=Ev.
\ee
If we treat the singularity in $S_{out}$ as above, we find an
imaginary part $-2\pi r_hE$, which leads to the correct Hawking
temperature, and the 
incoming solution involves no decay or amplification. 
The reason why the imaginary parts here differ from those
with Schwarzschild coordinates is that the transformation
between these coordinates involves the singular $r$ integral.

It is interesting to repeat the calculation with retarded
Eddington-Finkelstein coordinates. Here one uses
\bb
du=dt-{dr\over 1-\frac{r_h}{r}}
\ee
instead of $dv$ and the metric reads 
\bb
ds^2=-(1-\frac{r_h}{r})du^2- 2dudr+r^2d\Omega^2.
\ee
The Hamilton-Jacobi equation for spherically symmetric $S$ is
\bb
-2{\partial S\over \partial u}{\partial S\over \partial r}+
(1-\frac{r_h}{r})({\partial S\over \partial r})^2=0,
\ee
so that
\bb
S_{out}=Eu,\quad
S_{in}=Eu+2E\int{dr\over 1-\frac{r_h}{r}}.
\ee
Here, $S_{out}$ has no imaginary part but $S_{in}$ has an
imaginary part $2r_h\pi E$, which means that outward emission is
complete, but {\it absorption is inhibited} by a factor
$\exp (-2\pi r_hE/\hbar)$ in the amplitude,
and a factor $\exp (-4\pi r_hE/\hbar)$ in the probability.
We postpone the interpretation of this situation to the last section.

We now come to non-singular coordinates due to Painlev\'{e}, which were mentioned
in \cite{mitra}. There are two metrics of this kind:
\bb
ds^2=-(1-\frac{r_h}{r})dt^2\pm 2\sqrt{\frac{r_h}{r}}dtdr+dr^2+r^2d\Omega^2.
\ee
Only the upper sign in the cross term was used in the discussion given
in  \cite{mitra}. We shall extend the discussion to the lower sign and
find a difference. The Hamilton-Jacobi equation for spherically symmetric $S$ is
\bb
-({\partial S\over \partial t})^2\pm 2\sqrt{\frac{r_h}{r}}{\partial S\over \partial t}
{\partial S\over \partial r}+(1-\frac{r_h}{r})({\partial S\over \partial r})^2=0.
\ee
The $r$-dependent part of $S$ is then
\bb
S_0(r)= E\int^r{dr(\mp\sqrt{\frac{r_h}{r}}+(\pm 1))\over  1-\frac{r_h}{r}},
\ee
where the $\mp$ sign comes from the $\pm$ in the above equation, while
the  $(\pm 1)$  appears when the quadratic equation is solved. Thus, as in
the coordinates studied above,
\bb
S_{out}&=&Et-E[r_h\cdot i\pi(1\pm 1)+\nonumber\\
&&\int^r dr P(\frac{1}{r-r_h})r(1\pm \sqrt{\frac{r_h}{r}})],\nonumber\\
S_{in}&=&Et+E[r_h\cdot i\pi(1\mp 1)+\nonumber\\
&&\int^r dr P(\frac{1}{r-r_h})(1\mp \sqrt{\frac{r_h}{r}})],
\ee
For the upper metric, $Im~S_{out}=-2Er_h\pi,$
reproducing the Hawking temperature, while absorption is complete. 
When the lower sign is taken in the metric, emission is complete,
but absorption is inhibited by the factor 
$\exp (-2\pi r_hE/\hbar)$ in the amplitude,
and the factor $\exp (-4\pi r_hE/\hbar)$ in the probability.

Let us next look at Lema\^{i}tre coordinates for the same black hole -- cf. \cite{pad}.
Here, the metrics are (again one can choose either the + or the - sign)
\bb
ds^2=-d\tau^2+{dR^2\over [\frac{3}{2r_h}(R\mp\tau)]^{2/3}}
+r_h^2[\frac{3}{2r_h}(R\mp\tau)]^{4/3}d\Omega^2.
\ee
The Hamilton-Jacobi equation for spherically symmetric $S$ is
\bb
-({\partial S\over \partial\tau})^2 + [\frac{3}{2r_h}(R\mp\tau)]^{2/3}
({\partial S\over \partial R})^2=0.
\ee
We may write
\bb
S=S_-(R-\tau)+S_+(R+\tau).
\ee
For the upper sign, $S_+'=E/2$, where the normalization of the energy is fixed by
noting that $\tau=\frac12(R+\tau)-\frac12(R-\tau)$ and in the asymptotic region
the $dR^2$ piece of the metric drops out.
For the lower sign, the situation gets reversed: $S_-'=-E/2$.
The Hamilton-Jacobi equation thus reduces to
\bb
(\pm E/2+S_\mp')^2[\frac{3}{2r_h}(R\mp\tau)]^{2/3}=(E/2\mp S_\mp')^2,
\ee
so that
\bb
S=\pm\frac{E}{2}(R\pm\tau)+\frac{E}{2}(\frac{2r_h}{3})
\int d\xi_\mp^3{(\pm 1)\mp\xi_\mp\over \xi_\mp \pm (\pm 1)},
\ee
where $\xi_\mp\equiv[\frac{3}{2r_h}(R\mp\tau)]^{1/3}$.
The terms $(\pm 1)$ in the $\xi$ integral are unrelated to
the choice of sign in the metric and arise on solving
the quadratic equation for $S_\pm$.

The upper sign in  $(\pm 1)$ corresponds to incoming particles
and the lower sign corresponds to outgoing ones.
For the upper sign in the metric,
there is no singularity in the incoming case, and no imaginary part in 
the integral, while there is a singularity at the horizon
in the outgoing case, where the imaginary part of the integral comes out
to be $-2\pi Er_h$. 
For the lower sign in the metric, there is no singularity in the
outgoing case, and no imaginary part in the integral, while there is a 
singularity at the horizon
$\xi=1$ in the incoming case where the imaginary part of the integral comes out
to be $2\pi Er_h$. 
These observations lead to the expected decay factor
$\exp(-2\pi Er_h/\hbar)$ in the amplitude and
$\exp(-4\pi Er_h/\hbar)$ in the probability 
for emission with the upper metric, where absorption is complete.
The same factors hold also for absorption with
the lower metric, where emission is complete.

It may be instructive to note the differences of this treatment
with that in \cite{pad}. The normalization of $E$ is different.
Only one metric is used at a time, not both simultaneously. Lastly, it is not
necessary to invoke global factors to derive the Hawking temperature
for the upper sign of the metric.

One can also discuss the black hole in Kruskal coordinates. The metric is written as
\bb
ds^2=-{4r_h^3\over r}e^{-\frac{r}{r_h}}dUdV+r^2d\Omega^2,
\ee
where $r$ is related to the standard Kruskal coordinates by
\bb
UV=(1-\frac{r}{r_h})e^{\frac{r}{r_h}}.
\ee
The Hamilton-Jacobi equation satisfied by a spherically symmetric
$S$ is
\bb
{\partial S\over \partial U}{\partial S\over \partial V}=0.
\ee
There are two possibilities: $S$ is dependent on only $U$ (outgoing)
or only $V$ (incoming). These solutions take the form
\bb
S_{out}=-\frac{E}{\kappa}\int \frac{dU}{U},\quad
S_{in}=\frac{E}{\kappa}\int \frac{dV}{V}.
\ee
An imaginary piece arises in the
$U$ integral, and as in the above cases, it is $i\pi$. 
It should be pointed out here
that $U$ changes sign across the horizon, unlike proper radial variables,
which become imaginary on one side; thus an $i\pi$ factor arises for $U$,
just as for $r$, though it gets halved for proper radial variables.
This $i\pi$ is just what
is needed to reproduce the Hawking temperature $\frac{\hbar\kappa}{2\pi}$
for emission. 
Note that $S_{in}$ has a non-zero imaginary part $\frac{E\pi}{\kappa}$,
so that absorption is inhibited by the factor 
$\exp(-2\pi Er_h/\hbar)$ in the amplitude and
$\exp(-4\pi Er_h/\hbar)$ in the probability. 

Lastly we consider the Kerr black hole, which is 
a rotating one. It can be described by the metric
\bb
ds^2&=&-F(r,\theta)dt^2+{dr^2\over g(r,\theta)}+K(r,\theta)[d\phi
-\frac{H(r,\theta)}{K(r,\theta)}dt]^2\nonumber\\
&+&\Sigma(r,\theta)d\theta^2,
\ee
where
\bb
F(r,\theta)&=&{\Delta(r)\Sigma(r,\theta)\over (r^2+a^2)^2-\Delta(r)
a^2\sin^2\theta},\nonumber\\
g(r,\theta)&=&{\Delta(r)\over\Sigma(r,\theta)},\quad 
H(r,\theta)={2aMr\sin^2\theta\over \Sigma(r,\theta)},\nonumber\\
\Sigma(r,\theta)&=&r^2+a^2\cos^2\theta,\quad \Delta(r)=r^2+a^2-2Mr,\nonumber\\
K(r,\theta)&=&{(r^2+a^2)^2-\Delta(r)a^2\sin^2\theta\over \Sigma(r,\theta)}\sin^2\theta.
\ee
Here $M$ represents the mass and $a$ is a rotation parameter.
The Hamilton-Jacobi equation for $S(t,r,\phi)$ at a fixed $\theta=\theta_0$ 
near the horizon, which is located at
$r_+=M+\sqrt{M^2-a^2}$, satisfying $\Delta(r_+)=0,$ is
\bb
g(r,\theta_0)({\partial S\over\partial r})^2-\frac{1}{F(r,\theta_0)}
({\partial S\over\partial t}+\Omega{\partial S\over\partial\phi})^2=0,
\ee
where $\Omega\equiv\frac{H(r_+,\theta)}{K(r_+,\theta)}=\frac{a}{r_+^2+a^2}$.
Both $F(r,\theta_0)$ and $g(r,\theta_0)$ have simple zeroes at the horizon.
Near the horizon, $\sqrt{Fg}={\Delta\over r_+^2+a^2}$,
which is independent of $\theta_0$.
Discussions of rotating black holes are available in \cite{zerbini,mann}.
We shall however consider advanced Eddington-Finkelstein
coordinates defined by
\bb
dv=dt+dr {r^2+a^2\over \Delta},\quad d\psi=d\phi+dr{a\over \Delta}.\ee
Then it is straightforward to see that near the horizon, $S(v,r,\psi)$ satisfies
\bb
({\partial S\over\partial r}+{\partial S\over\partial v}
\frac{r_+^2+a^2}{\Delta}+{\partial S\over\partial \psi}{a\over \Delta})^2&=&
(\frac{r_+^2+a^2}{\Delta})^2({\partial S\over\partial v}\nonumber\\
&+&\Omega{\partial S\over\partial\psi})^2.
\ee
There are two solutions:
\bb
S_{out}&=&Ev+J\psi-2(E+\Omega J)\int{dr (r_+^2+a^2)\over \Delta},\nonumber\\
S_{in}&=&Ev+J\psi. 
\ee
The incoming solution shows that absorption is complete, while the
outgoing solution has an imaginary part and involves
a decay factor of $\exp[-(E+\Omega J)\pi(r_+^2+a^2)/\hbar(r_+-M)]$ in
the amplitude and hence $\exp[-2(E+\Omega J)\pi(r_+^2+a^2)/\hbar(r_+-M)]$
in the probability, indicating the known temperature 
${\hbar(r_+-M)\over 2\pi(r_+^2+a^2)}$.

The calculation may be repeated with retarded Eddington-Finkelstein
coordinates defined by
\bb
du=dt-dr {r^2+a^2\over \Delta},\quad d\chi=d\phi-dr{a\over \Delta},
\ee
leading to
\bb
S_{out}&=&Eu+J\chi,\nonumber\\
S_{in}&=&Eu+J\chi+2(E+\Omega J)\int{dr (r_+^2+a^2)\over \Delta}.
\ee
In this case the outgoing solution has no imaginary part and shows
complete emission while the incoming solution has an imaginary part
leading to the same decay factor as above. 


\noindent{\bf 3.} In this letter we have sought to analyse tunnelling across
horizons in the Hamilton-Jacobi approach using non-singular
coordinates and forsaking the imaginary part of the integration constant $C$ 
that was found to be necessary in other coordinates \cite{mitra}.
In the preceding sections we have observed that different non-singular coordinate
systems lead to two kinds of results. For one set, the incoming
solutions do not contain any imaginary parts, implying complete absorption.
Correspondingly, the outgoing solutions contain an imaginary part which
is just right to support the interpretation of the inhibition
associated with the Hawking temperature. This is comforting.
The advanced Eddington-Finkelstein coordinates, the upper Painlev\'{e}
and the upper Lema\^{i}tre coordinates belong to this category.

For another set of non-singular coordinates, the situation is just the 
reverse. The outgoing solution has no imaginary part, suggesting
that emission is uninhibited, while the incoming solution has just the
right imaginary part to suggest an inhibition of the kind which
would be expected with the Hawking temperature.
The retarded Eddington-Finkelstein coordinates, the lower Painlev\'{e}
and the lower Lema\^{i}tre coordinates belong to this category.
There is also the Kruskal choice of coordinates where both the
outgoing and the incoming solutions contain imaginary parts,
corresponding to the kind of inhibition seen above. What can one make of
this?

One has to remember that while the advanced Eddington-Finkelstein coordinates, 
the upper Painlev\'{e} and the upper Lema\^{i}tre coordinates all describe
black holes with their exterior and interior regions, the retarded 
Eddington-Finkelstein coordinates, the lower Painlev\'{e}
and the lower Lema\^{i}tre coordinates actually describe their time-reversed
versions, {\it i.e.,} the exterior regions with white hole interiors. 
Just as black holes may absorb matter classically
but can emit matter only quantum mechanically and 
with a specific temperature, white holes may emit matter classically but can absorb 
matter only quantum mechanically and again with a 
specific temperature. This is the meaning of the imaginary parts found above. 
The Kruskal case is different: here one simultaneously has a black hole region 
and a white hole region. However, if we consider the black hole and the 
exterior regions only, which is equivalent to restricting the $V$-coordinate 
to the domain $0\leq V\leq\infty$, we effectively recover the 
results obtained in the advanced Eddington-Finkelstein coordinates. Similar 
conclusions hold when we consider the white hole and the exterior regions only.

White holes are perhaps less familiar objects than black holes, but
they are essentially time-reversed versions of the latter.
Hawking has argued \cite{hawk} that while black holes may be created
by classical collapse, their disintegration is a quantum phenomenon.
By time-reversal, the creation of white holes is a quantum phenomenon,
but their disintegration is classical. This is consistent with our
results, though of course we consider only eternal black or white
holes, which are used only as providing gravitational backgrounds. 

As regards the integration constant $C$, we have dispensed with it
in discussions using non-singular coordinates, but it is possible for an imaginary part 
of $C$ to appear for Schwarzschild and other coordinates which are not 
well-defined across the horizon. 
Distinct boundary conditions (imaginary part of $C$) 
are to be imposed to fix whether a black or a white hole is being considered.
Only by insisting that absorption is uninhibited can we be sure of getting a
black hole and the correct emission probability. Similarly, a boundary condition
of unhindered emission will ensure the identification of a white hole and its
absorption probability. These boundary conditions can be tuned for singular
coordinates by allowing $C$ to have an imaginary part, which 
corresponds to an imaginary part in the transformation taking
a non-singular system of coordinates to another system. Similar boundary conditions
for non-singular coordinates lead to ${\rm Im}\,C=0$ as is implicit in our calculations.

\end{document}